\documentclass[twoside,reqno]{lt8-proc}
\usepackage{epsfig,cite}
\usepackage{amssymb,amsmath}
\usepackage{times}
\newcommand\rf[1]{(\ref{eq:#1})}
\newcommand\lab[1]{\label{eq:#1}}
\newcommand\nonu{\nonumber}
\newcommand\br{\begin{eqnarray}}
\newcommand\er{\end{eqnarray}}
\newcommand\be{\begin{equation}}
\newcommand\ee{\end{equation}}

\newcommand\foot[1]{\footnotemark\footnotetext{#1}}
\newcommand\lb{\lbrack}
\newcommand\rb{\rbrack}

\newcommand\llb{\left\lbrack}
\newcommand\rrb{\right\rbrack}

\newcommand\lcurl{\left\{}
\newcommand\rcurl{\right\}}
\renewcommand\({\left(}
\renewcommand\){\right)}
\newcommand\bv{\bigm\vert}               
\newcommand\bgv{\bigg\vert}              

\newcommand\bc{\begin{center}}
\newcommand\ec{\end{center}}

\newcommand\partder[2]{\frac{{\partial {#1}}}{{\partial {#2}}}}

\renewcommand\b{\beta}

\renewcommand\d{\delta}

\newcommand\eps{\epsilon}
\newcommand\vareps{\varepsilon}
\newcommand\g{\gamma}
\newcommand\G{\Gamma}

\newcommand\h{\frac{1}{2}}
\renewcommand\k{\kappa}
\renewcommand\l{\lambda}
\renewcommand\L{\Lambda}
\newcommand\m{\mu}
\newcommand\n{\nu}

\newcommand\vp{\varphi}
\renewcommand\P{\Phi}
\newcommand\pa{\partial}

\newcommand\pr{\prime}

\renewcommand\r{\rho}
\newcommand\s{\sigma}

\renewcommand\t{\tau}
\renewcommand\th{\theta}

\newcommand\wti{\widetilde}

\newcommand\cA{{\mathcal A}}

\newcommand\cF{{\mathcal F}}

\newcommand{\ct}[1]{\cite{#1}}
\newcommand{\bib}[1]{\bibitem{#1}}
\newcommand\PRL[3]{\textsl{Phys. Rev. Lett.} \textbf{#1}, #3 (#2)}
\newcommand\NPB[3]{\textsl{Nucl. Phys.} \textbf{B#1}, #3 (#2)}

\newcommand\PRD[3]{\textsl{Phys. Rev.} \textbf{D#1}, #3 (#2)}

\newcommand\PLB[3]{\textsl{Phys. Lett.} \textbf{#1B}, #3 (#2)}
\newcommand\CQG[3]{\textsl{Class. Quantum Grav.} \textbf{#1}, #3 (#2)}

\newcommand\AoP[3]{\textsl{Ann. of Phys.} \textbf{#1}, #3 (#2)}

\newcommand\IJMPA[3]{\textsl{Int. J. Mod. Phys.} \textbf{A#1}, #3 (#2)}

\newcommand\Xdot{\stackrel{.}{X}}

\newcommand\rdot{\stackrel{.}{r}}

\begin{document}
\sloppy \raggedbottom
\setcounter{page}{1}

\newpage
\setcounter{figure}{0}
\setcounter{equation}{0}
\setcounter{footnote}{0}
\setcounter{table}{0}
\setcounter{section}{0}



\title{Asymmetric Wormholes via Electrically Charged Lightlike Branes}

\runningheads{Guendelman, Kaganovich, Nissimov and Pacheva}{Asymmetric Wormholes 
via Electrically Charged Lightlike Branes}

\begin{start}


\coauthor{E. Guendelman\thanks{guendel@bgu.ac.il}}{1},
\coauthor{A. Kaganovich\thanks{alexk@bgu.ac.il}}{1},
\author{E. Nissimov\thanks{nissimov@inrne.bas.bg}}{2},\\
\coauthor{S. Pacheva\thanks{nissimov@inrne.bas.bg, svetlana@inrne.bas.bg}}{2}

\address{Department of Physics, Ben-Gurion University of the Negev, P.O.Box 653,\\ 
IL-84105 ~Beer-Sheva, Israel}{1}
\address{Institute for Nuclear Research and Nuclear Energy, Bulgarian Academy of \\
Sciences, Boul. Tsarigradsko Chausee 72, BG-1784 ~Sofia, Bulgaria}{2}


\begin{Abstract}
We consider a self-consistent Einstein-Maxwell-Kalb-Ramond system in the bulk $D=4$ 
space-time interacting with a variable-tension electrically charged {\em lightlike}
brane. The latter serves both as a material and charge source for gravity
and electromagnetism, as well as it {\em dynamically} generates a bulk space varying 
cosmological constant. We find an {\em asymmetric wormhole} solution describing
two ``universes'' with different spherically symmetric black-hole-type geometries
connected through a ``throat'' occupied by the lightlike brane. The electrically 
neutral ``left universe'' comprises the exterior region of Schwarzschild-de-Sitter
(or pure Schwarzschild) space-time above the {\em inner} (Schwarzschild-type) horizon,
whereas the electrically charged ``right universe'' consists of the exterior 
Reissner-Nordstr{\"o}m (or Reissner-Nordstr{\"o}m-de-Sitter) black hole region beyond
the {\em outer} Reissner-Nordstr{\"o}m horizon. All physical parameters of the 
wormhole are uniquely determined by two free parameters -- the electric charge and 
Kalb-Ramond coupling of the lightlike brane.
\end{Abstract}
\end{start}


\section{Introduction}

Lightlike brane (\textsl{LL-branes} for short) 
play an important role in general relativity as they enter the description of
various physically important cosmological and astrophysical phenomena such as: 
(i) impulsive lightlike signals arising in cataclysmic astrophysical events 
\ct{barrabes-hogan}; (ii) the ``membrane paradigm'' \ct{membrane-paradigm} of black 
hole physics; (iii) the thin-wall approach to domain walls coupled to 
gravity \ct{Israel-66,Barrabes-Israel,Dray-Hooft}.
More recently, \textsl{LL-branes} became significant also in the context of
modern non-perturbative string theory, in particular, as the so called
$H$-branes describing quantum horizons (black hole and cosmological)
\ct{kogan-01}, as Penrose limits of baryonic $D$-branes
\ct{mateos-02}, \textsl{etc} (see also Refs.\ct{nonperturb-string}).

In the pioneering papers \ct{Israel-66,Barrabes-Israel,Dray-Hooft} \textsl{LL-branes}
in the context of gravity and cosmology have been extensively studied from a 
phenomenological point of view, \textsl{i.e.}, by introducing them without specifying
the Lagrangian dynamics from which they may originate\foot{In a more recent paper 
\ct{barrabes-israel-05} brane actions in terms of their pertinent extrinsic geometry
have been proposed which generically describe non-lightlike branes, whereas the 
lightlike branes are treated as a limiting case.}. 
On the other hand, we have proposed in a series of recent papers 
\ct{LL-brane-main,inflation-all,our-WH,rot-WH} a new class of concise Lagrangian actions, 
providing a derivation from first principles of the \textsl{LL-brane} dynamics.

There are several characteristic features of \textsl{LL-branes} which drastically
distinguish them from ordinary Nambu-Goto branes: 

(i) They describe intrinsically lightlike modes, whereas Nambu-Goto branes describe
massive ones.

(ii) The tension of the \textsl{LL-brane} arises as an {\em additional
dynamical degree of freedom}, whereas Nambu-Goto brane tension is a given
{\em ad hoc} constant. 
The latter characteristic feature significantly distinguishes our \textsl{LL-brane}
models from the previously proposed {\em tensionless} $p$-branes (for a review, 
see Ref.\ct{lindstroem-etal}) which rather resemble a $p$-dimensional continuous
distribution of massless point-particles. 

(iii) Consistency of \textsl{LL-brane} dynamics in a spherically or axially
symmetric gravitational background of codimension one requires the presence
of an event horizon which is automatically occupied by the \textsl{LL-brane}
(``horizon straddling'' according to the terminology of 
Ref.\ct{Barrabes-Israel}).

(iv) When the \textsl{LL-brane} moves as a {\em test} brane in spherically or 
axially symmetric gravitational backgrounds its dynamical tension exhibits 
exponential ``inflation/deflation'' time behaviour \ct{inflation-all}
-- an effect similar to the ``mass inflation'' effect around black hole horizons
\ct{israel-poisson}. 

In a series of recent papers \ct{our-WH,rot-WH,ER-bridge} we have explored the novel 
possibility of employing \textsl{LL-branes} as natural self-consistent gravitational
sources for wormhole space-times, in other words, generating wormhole solutions in
self-consistent bulk gravity-matter systems coupled to
\textsl{LL-branes} through dynamically derived world-volume \textsl{LL-brane}
stress energy tensors. For a review of wormhole space-times, 
see Refs.\ct{visser-book,WH-rev}.

The possibility of a ``wormhole space-time'' was first hinted at in the work
of Einstein and Rosen \ct{einstein-rosen}, where they considered matching at
the horizon of two identical copies of the exterior Schwarzschild space-time
region (subsequently called {\em Einstein-Rosen ``bridge''}).
The original Einstein-Rosen ``bridge'' manifold appears as a particular case of the
construction of spherically symmetric wormholes produced by
\textsl{LL-branes} as gravitational sources 
(Refs.\ct{ER-bridge,rot-WH}; see also Section 5 below). 
The main lesson here is that consistency of Einstein
equations of motion yielding the original Einstein-Rosen ``bridge'' as well-defined
solution necessarily requires the presence of \textsl{LL-brane}
energy-momentum tensor as a source on the right hand side.
Thus, the introduction of \textsl{LL-brane} coupling to gravity 
brings the original Einstein-Rosen construction in Ref.\ct{einstein-rosen} 
to a consistent completion\foot{Let us particularly emphasize that here and in what 
follows we consider the Einstein-Rosen ``bridge'' in  its original formulation in 
Ref.\ct{einstein-rosen} as a four-dimensional space-time manifold consisting of two 
copies of the exterior Schwarzschild space-time region matched along the horizon.
On the other hand, the nomenclature of ``Einstein-Rosen bridge'' in several standard 
textbooks (\textsl{e.g.} Ref.\ct{MTW}) 
uses the Kruskal-Szekeres manifold. The latter notion of ``Einstein-Rosen bridge''
is not equivalent to the original construction in Ref.\ct{einstein-rosen}. Namely, 
the two regions in Kruskal-Szekeres space-time corresponding to the outer 
Schwarzschild space-time region ($r>2m$) and labeled $(I)$ and $(III)$ 
in Refs.\ct{MTW} are generally
{\em disconnected} and share only a two-sphere (the angular part) as a common border
($U=0, V=0$ in Kruskal-Szekeres coordinates), whereas in the original Einstein-Rosen
``bridge'' construction the boundary between the two identical copies of the
outer Schwarzschild space-time region ($r>2m$) is a three-dimensional hypersurface
($r=2m)$.}.

More complicated examples of spherically and axially symmetric wormholes with
Reissner-Nordstr{\"o}m and rotating cylindrical geometry, respectively,
have also been presented in Refs.\ct{our-WH,rot-WH}. Namely,
two copies of the outer space-time region of a Reissner-Nordstr{\"o}m or
rotating cylindrical black hole, respectively, 
are matched via \textsl{LL-brane} along what used to be the outer horizon of the
respective full black hole space-time manifold. In this way we
obtain a wormhole solution which combines the features of the Einstein-Rosen
``bridge'' on the one hand (with wormhole throat at horizon), and the features of
Misner-Wheeler wormholes \ct{misner-wheeler}, \textsl{i.e.}, exhibiting the so called 
``charge without charge'' phenomenon\foot{Misner and Wheeler \ct{misner-wheeler}
realized that wormholes connecting two asymptotically flat space times provide 
the possibility of ``charge without charge'', \textsl{i.e.}, electromagnetically
non-trivial solutions where the lines of force of the electric field flow from one 
universe to the other without a source and giving the impression of being 
positively charged in one universe and negatively charged in the other universe.},
on the other hand.

In the present note the results of Refs.\ct{our-WH,rot-WH} will be 
extended to the case of {\em asymmetric} wormholes, describing two
``universes'' with different spherically symmetric geometries of black hole
type connected via a ``throat'' occupied by the pertinent gravitational
source -- an electrically charged \textsl{LL-brane}. As a result of the
well-defined world-volume \textsl{LL-brane} dynamics coupled
self-consistently to gravity and bulk space-time gauge fields, it creates a
``left universe'' with Schwarzschild-de-Sitter geometry where the
cosmological constant is dynamically generated, and a ``right universe''
with Reissner-Nordstr{\"o}m geometry with dynamically generated Coulomb
field-strength. Similarly, the \textsl{LL-brane} can dynamically generate a non-zero
cosmological constant in the ``right universe'', in which case it connects a
purely Schwarzschild ``left universe'' with a Reissner-Nordstr{\"o}m-de-Sitter
``right universe''.

The presentation of the material goes as follows. In Section 2 we briefly review the
reparametrization-invariant world-volume Lagrangian formulation of \textsl{LL-branes}
in both the Polyakov-type and Nambu-Goto-type forms. In Section 3 we briefly describe 
the main properties of \textsl{LL-brane} dynamics in spherically symmetric
gravitational backgrounds stressing particularly on the ``horizon straddling'' 
phenomenon and the dynamical cosmological constant generation. Section 4
contains our principal result -- the explicit construction of an asymmetric
wormhole solution of self-consistent Einstein-Maxwell-Kalb-Ramond system
interacting with an electrically charged \textsl{LL-brane}. The wormhole space-time 
consists of two ``universes'' with different spherically symmetric geometries
connected through a ``throat'' occupied by the \textsl{LL-brane}: (a) electrically 
neutral ``left universe'' comprising the exterior region of Schwarzschild-de-Sitter
(or pure Schwarzschild) space-time above the {\em inner} (Schwarzschild-type) horizon;
(b) electrically charged ``right universe'' consisting of the exterior 
Reissner-Nordstr{\"o}m (or Reissner-Nordstr{\"o}m-de-Sitter)
black hole region beyond the {\em outer} Reissner-Nordstr{\"o}m horizon. 
In Section 5 we briefly consider the simple special case
of the above construction with vanishing charge and Kalb-Ramond coupling of
the \textsl{LL-branes}. It consistently describes the famous Einstein-Rosen
``bridge'' wormhole solution \ct{einstein-rosen}. On the way we explain the crucial 
role of the presence of the \textsl{LL-brane} gravitational source producing the Einstein-Rosen ``bridge'' -- an observation missing in the original classic paper \ct{einstein-rosen}.


\section{Einstein-Maxwell-Kalb-Ramond System Interacting With Lightlike Brane:
Lagrangian Formulation}


Self-consistent bulk Einstein-Maxwell-Kalb-Ramond system coupled to a charged 
codimension-one {\em lightlike} $p$-brane (\textsl{i.e.},
$D=(p+1)+1$) is described by the following action:
\be
S = \int\!\! d^D x\,\sqrt{-G}\,\llb \frac{R(G)}{16\pi} 
- \frac{1}{4} \cF_{\m\n}\cF^{\m\n} 
- \frac{1}{D! 2} \cF_{\m_1\ldots\m_D}\cF^{\m_1\ldots\m_D}\rrb 
+ {\wti S}_{\mathrm{LL}} \; .
\lab{E-M-KR+LL}
\ee
Here $\cF_{\m\n} = \pa_\m \cA_\n - \pa_\n \cA_\m$ and 
\be
\cF_{\m_1\ldots\m_D} = D\pa_{[\m_1} \cA_{\m_2\ldots\m_D]} =
\cF \sqrt{-G} \vareps_{\m_1\ldots\m_D}
\lab{F-KR}
\ee
are the field-strengths of the electromagnetic $\cA_\m$ and Kalb-Ramond 
$\cA_{\m_1\ldots\m_{D-1}}$ gauge potentials \ct{aurilia-townsend}.
The last term on the r.h.s. of \rf{E-M-KR+LL} indicates the reparametrization
invariant world-volume action of the \textsl{LL-brane} coupled to the bulk
gauge fields, proposed in our previous papers 
\ct{LL-brane-main,inflation-all,our-WH}:
\br
{\wti S}_{\mathrm{LL}} = S_{\mathrm{LL}}
- q \int d^{p+1}\s\,\vareps^{ab_1\ldots b_p} F_{b_1\ldots b_p} \pa_a X^\m \cA_\m
\nonu \\
- \frac{\b}{(p+1)!} \int d^{p+1}\s\,\vareps^{a_1\ldots a_{p+1}}
\pa_{a_1} X^{\m_1}\ldots\pa_{a_{p+1}} X^{\m_{p+1}} \cA_{\m_1\ldots\m_{p+1}} \; .
\lab{LL-action+EM+KR}
\er
where:
\be
S_{\mathrm{LL}} = \int d^{p+1}\s\,\P\llb -\h\g^{ab} g_{ab} + L\!\( F^2\)\rrb \; .
\lab{LL-action}
\ee
In Eqs.\rf{LL-action+EM+KR}--\rf{LL-action} the following notions and notations 
are used:

\begin{itemize}
\item
$\P$ is alternative non-Riemannian integration measure density (volume form) 
on the $p$-brane world-volume manifold:
\br
\P \equiv \frac{1}{(p+1)!} 
\vareps^{a_1\ldots a_{p+1}} H_{a_1\ldots a_{p+1}}(B) \;\; ,
\lab{mod-measure-p} \\
H_{a_1\ldots a_{p+1}}(B) = (p+1) \pa_{[a_1} B_{a_2\ldots a_{p+1}]} \; ,
\lab{H-def}
\er
instead of the usual $\sqrt{-\g}$. Here $\vareps^{a_1\ldots a_{p+1}}$ is the
alternating symbol ($\vareps^{0 1\ldots p} = 1$), $\g_{ab}$ ($a,b=0,1,{\ldots},p$)
indicates the intrinsic Riemannian metric on the world-volume, and
$\g = \det\Vert\g_{ab}\Vert$.
$H_{a_1\ldots a_{p+1}}(B)$ denotes the field-strength of an auxiliary
world-volume antisymmetric tensor gauge field $B_{a_1\ldots a_{p}}$ of rank $p$.
As a special case one can build $H_{a_1\ldots a_{p+1}}$ \rf{H-def} in terms of 
$p+1$ auxiliary world-volume scalar fields $\lcurl \vp^I \rcurl_{I=1}^{p+1}$:
\be
H_{a_1\ldots a_{p+1}} = \vareps_{I_1\ldots I_{p+1}}
\pa_{a_1} \vp^{I_1}\ldots \pa_{a_{p+1}} \vp^{I_{p+1}} \;.
\lab{mod-measure-p-scalar}
\ee
Note that $\g_{ab}$ is {\em independent} of the auxiliary world-volume fields
$B_{a_1\ldots a_{p}}$ or $\vp^I$.
The alternative non-Riemannian volume form \rf{mod-measure-p}
has been first introduced in the context of modified standard (non-lightlike) string and
$p$-brane models in Refs.\ct{mod-measure}.
\item
$X^\m (\s)$ are the $p$-brane embedding coordinates in the bulk
$D$-dimensional space time with bulk Riemannian metric
$G_{\m\n}(X)$ with $\m,\n = 0,1,\ldots ,D-1$; 
$(\s)\equiv \(\s^0 \equiv \t,\s^i\)$ with $i=1,\ldots ,p$;
$\pa_a \equiv \partder{}{\s^a}$.
\item
$g_{ab}$ is the induced metric on world-volume:
\be
g_{ab} \equiv \pa_a X^{\m} \pa_b X^{\n} G_{\m\n}(X) \; ,
\lab{ind-metric}
\ee
which becomes {\em singular} on-shell (manifestation of the lightlike nature, 
cf. second Eq.\rf{on-shell-singular} below).
\item
$L\!\( F^2\)$ is the Lagrangian density of another
auxiliary $(p-1)$-rank antisymmetric tensor gauge field $A_{a_1\ldots a_{p-1}}$
on the world-volume with $p$-rank field-strength and its dual:
\be
F_{a_1 \ldots a_{p}} = p \pa_{[a_1} A_{a_2\ldots a_{p}]} \quad ,\quad
F^{\ast a} = \frac{1}{p!} \frac{\vareps^{a a_1\ldots a_p}}{\sqrt{-\g}}
F_{a_1 \ldots a_{p}}  \; .
\lab{p-rank}
\ee
$L\!\( F^2\)$ is {\em arbitrary} function of $F^2$ with the short-hand notation:
\be
F^2 \equiv F_{a_1 \ldots a_{p}} F_{b_1 \ldots b_{p}} 
\g^{a_1 b_1} \ldots \g^{a_p b_p} \; .
\lab{F2-id}
\ee
\end{itemize}


Rewriting the action \rf{LL-action} in the following equivalent form:
\be
S = - \int d^{p+1}\!\s \,\chi \sqrt{-\g}
\Bigl\lb \h \g^{ab} \pa_a X^{\m} \pa_b X^{\n} G_{\m\n}(X) - L\!\( F^2\)\Bigr\rb
\;\; , \; \chi \equiv \frac{\P}{\sqrt{-\g}}
\lab{LL-action-chi}
\ee
with $\P$ the same as in \rf{mod-measure-p},
we find that the composite field $\chi$ plays the role of a {\em dynamical
(variable) brane tension}. The notion of dynamical brane tension has previously 
appeared in different contexts in Refs.\ct{townsend-etal}.

Let us also remark that, as it has been shown in Refs.\ct{our-WH,reg-BH}, the
\textsl{LL-brane} equations of motion corresponding to the Polyakov-type action 
\rf{LL-action} (or \rf{LL-action-chi}) can be equivalently obtained from the 
following {\em dual} Nambu-Goto-type action:
\br
S_{\rm NG} = - \int d^{p+1}\s \, T 
\sqrt{\bgv\,\det\Vert g_{ab} - \eps \frac{1}{T^2}\pa_a u \pa_b u\Vert\,\bgv} 
\quad ,\quad \eps = \pm 1 \; .
\lab{LL-action-NG}
\er
Here $T$ is {\em dynamical} tension simply proportional to the dynamical
tension in the Polyakov-type formulation \rf{LL-action} 
~($T\sim \chi = \frac{\P}{\sqrt{-\g}}$), and $u$ denotes the dual potential 
w.r.t. $A_{a_1\ldots a_{p-1}}$:
\be
F^{\ast}_{a} (A) = \mathrm{const}\, \frac{1}{\chi} \pa_a u \; .
\lab{u-def}
\ee
It what follows we will consider the original Polyakov-type action \rf{LL-action}.

The pertinent Einstein-Maxwell-Kalb-Ramond equations of motion derived from
the action \rf{E-M-KR+LL} read:
\be
R_{\m\n} - \h G_{\m\n} R =
8\pi \( T^{(EM)}_{\m\n} + T^{(KR)}_{\m\n} + T^{(brane)}_{\m\n}\) \; ,
\lab{Einstein-eqs}
\ee
\be
\pa_\n \(\sqrt{-G}\cF^{\m\n}\) + 
q \int\!\! d^{p+1}\s\,\d^{(D)}\Bigl(x-X(\s)\Bigr)
\vareps^{ab_1\ldots b_p} F_{b_1\ldots b_p} \pa_a X^\m = 0 \; ,
\lab{Maxwell-eqs}
\ee
\br
\vareps^{\n\m_1\ldots\m_{p+1}} \pa_\n \cF - 
\nonu \\
\b\,\int\! d^{p+1}\s\,\d^{(D)}(x - X(\s))
\vareps^{a_1\ldots a_{p+1}} \pa_{a_1}X^{\m_1}\ldots\pa_{a_{p+1}}X^{\m_{p+1}} = 0 \; ,
\lab{F-KR-eqs}
\er
where in the last equation we have used relation \rf{F-KR}. 
The explicit form of the energy-momentum tensors read:
\be
T^{(EM)}_{\m\n} = \cF_{\m\k}\cF_{\n\l} G^{\k\l} - G_{\m\n}\frac{1}{4}
\cF_{\r\k}\cF_{\s\l} G^{\r\s}G^{\k\l} \; ,
\lab{T-EM}
\ee
\br
T^{(KR)}_{\m\n} = \frac{1}{(D-1)!}\llb \cF_{\m\l_1\ldots\l_{D-1}}
{\cF_{\n}}^{\l_1\ldots\l_{D-1}} -
\frac{1}{2D} G_{\m\n} \cF_{\l_1\ldots\l_D} \cF^{\l_1\ldots\l_D}\rrb
\nonu \\
= - \h \cF^2 G_{\m\n} 
\lab{T-KR}
\er
\be
T^{(brane)}_{\m\n} = - G_{\m\k}G_{\n\l}
\int\!\! d^{p+1}\s\,\frac{\d^{(D)}\Bigl(x-X(\s)\Bigr)}{\sqrt{-G}}\,
\chi\,\sqrt{-\g} \g^{ab}\pa_a X^\k \pa_b X^\l \; ,
\lab{T-brane}
\ee
where the brane stress-energy tensor is straightforwardly derived
from the world-volume action \rf{LL-action} (or, equivalently, \rf{LL-action-chi};
recall $\chi\equiv\frac{\P}{\sqrt{-\g}}$ is the variable brane tension).

Eqs.\rf{Maxwell-eqs}--\rf{F-KR-eqs} show that:

(i) the \textsl{LL-brane} is charged source for the bulk electromagnetism;

(ii) the \textsl{LL-brane} uniquely determines the
value of $\cF^2$ in Eq.\rf{T-KR} through its coupling to the bulk
Kalb-Ramond gauge field (Eq.\rf{F-KR-eqs}) which implies {\em dynamical generation} of
bulk cosmological constant $\L=4\pi \cF^2$.


\section{Lightlike Brane Dynamics in Spherically Symmetric Gravitational Backgrounds}

The equations of motion of the \textsl{LL-brane} are discussed at length in our
previous papers \ct{LL-brane-main,inflation-all,our-WH,rot-WH}. Their explicit form
reads:
\be
\pa_a \Bigl\lb \h \g^{cd} g_{cd} - L(F^2)\Bigr\rb = 0 \quad \longrightarrow \quad
\h \g^{cd} g_{cd} - L(F^2) = M  \; ,
\lab{phi-eqs}
\ee
where $M$ is an arbitrary integration constant;
\be
\h g_{ab} - F^2 L^{\pr}(F^2) \llb\g_{ab} 
- \frac{F^{*}_a F^{*}_b}{F^{*\, 2}}\rrb = 0  \; ,
\lab{gamma-eqs}
\ee
where $F^{*\, a}$ is the dual world-volume field strength \rf{p-rank};
\br
\pa_{[a}\( F^{\ast}_{b]}\, \chi L^\pr (F^2)\) 
+ \frac{q}{4}\pa_a X^\m \pa_b X^\n \cF_{\m\n} = 0  \; ;
\lab{A-eqs} \\
\pa_a \(\chi \sqrt{-\g} \g^{ab} \pa_b X^\m\) + 
\chi \sqrt{-\g} \g^{ab} \pa_a X^\n \pa_b X^\l \G^\m_{\n\l}
\nonu \\
-q \vareps^{ab_1\ldots b_p} F_{b_1\ldots b_p} \pa_a X^\n \cF_{\l\n}G^{\l\m}
\nonu \\
- \frac{\b}{(p+1)!} \vareps^{a_1\ldots a_{p+1}} \pa_{a_1} X^{\m_1} \ldots
\pa_{a_{p+1}} X^{\m_{p+1}} \cF_{\l\m_1\dots\m_{p+1}} G^{\l\m} = 0 \; .
\lab{X-eqs}
\er
Here $\chi$ is the dynamical brane tension as in \rf{LL-action-chi}, 
$\cF_{\l\m_1\dots\m_{p+1}}$ is the Kalb-Ramond field-strength \rf{F-KR},
\be
\G^\m_{\n\l}=\h G^{\m\k}\(\pa_\n G_{\k\l}+\pa_\l G_{\k\n}-\pa_\k G_{\n\l}\)
\lab{affine-conn}
\ee
is the Christoffel connection for the external metric,
and $L^\pr(F^2)$ denotes derivative of $L(F^2)$ w.r.t. the argument $F^2$.

Eqs.\rf{phi-eqs}--\rf{gamma-eqs} imply the following important consequences:
\be
F^2 = F^2 (M) = \mathrm{const} \quad ,\quad g_{ab}F^{*\, b}=0 \; ,
\lab{on-shell-singular}
\ee
where the second equation is the manifestation of the lightlike nature of
the $p$-brane model \rf{LL-action}, \textsl{i.e.}, the tangent vector to the world-volume 
$F^{*\, a}\pa_a X^\m$ is {\em lightlike} w.r.t. metric of the embedding space-time.

World-volume reparametrization invariance allows us to introduce the standard 
synchronous gauge-fixing conditions:
\be
\g^{0i} = 0 \;\; (i=1,\ldots,p) \; ,\; \g^{00} = -1 \; .
\lab{gauge-fix}
\ee
Also, we will use a natural ansatz for the ``electric'' part of the 
auxiliary world-volume gauge field-strength \rf{p-rank}:
\be
F^{\ast i}= 0 \;\; (i=1,{\ldots},p) \quad ,\quad \mathrm{i.e.} \;\;
F_{0 i_1 \ldots i_{p-1}} = 0 \; ,
\lab{F-ansatz}
\ee
meaning that we choose the lightlike direction in Eq.\rf{on-shell-singular} 
to coincide with the brane
proper-time direction on the world-volume ($F^{*\, a}\pa_a \sim \pa_\t$).
The Bianchi identity ($\nabla_a F^{\ast\, a}=0$) together with 
\rf{gauge-fix}--\rf{F-ansatz} and the definition for the dual field-strength
in \rf{p-rank} imply:
\be
\pa_0 \g^{(p)} = 0 \quad \mathrm{where}\;\; \g^{(p)} \equiv \det\Vert\g_{ij}\Vert \; .
\lab{gamma-p-0}
\ee

Taking into account \rf{gauge-fix}--\rf{F-ansatz}, Eqs.\rf{gamma-eqs}
acquire the following gauge-fixed form (recall definition of the induced metric
$g_{ab}$ \rf{ind-metric}):
\be
g_{00}\equiv \Xdot^\m\!\! G_{\m\n}\!\! \Xdot^\n = 0 \quad ,\quad g_{0i} = 0 \quad ,\quad
g_{ij} - 2a_0\, \g_{ij} = 0 \; ,
\lab{gamma-eqs-0}
\ee
where $a_0$ is a $M$-dependent constant:
\be
a_0 \equiv F^2 L^\pr (F^2)\bv_{F^2 = F^2(M)} \; .
\lab{a0-const}
\ee
Eqs.\rf{gamma-eqs-0} are analogs of the Virasoro constraints in standard string theory.

In what follows we will be interested in static spherically symmetric solutions 
of Einstein-Maxwell-Kalb-Ramond equations \rf{Einstein-eqs}--\rf{F-KR-eqs}.
The generic form of spherically symmetric metric in Eddington-Finkelstein 
coordinates \ct{EFM} reads:
\be
ds^2 = - A(r) dv^2 + 2 dv\,dr + C(r) h_{ij}(\th) d\th^i d\th^j \; ,
\lab{EF-metric}
\ee
where $h_{ij}$ indicates the standard metric on $S^p$.
We will consider the simplest ansatz for the \textsl{LL-brane} embedding
coordinates:
\be
X^0\equiv v = \t \quad, \quad X^1\equiv r = r(\t) \quad, \quad 
X^i\equiv \th^i = \s^i \;\; (i=1,\ldots ,p)
\lab{X-embed}
\ee
Now, the \textsl{LL-brane} equations \rf{gamma-eqs-0} together with \rf{gamma-p-0}
yield:
\be
-A(r) + 2\rdot = 0 \quad , \quad \pa_\t C = \rdot\,\pa_r C\bv_{r=r(\t)} = 0 \; ,
\lab{r-const}
\ee
implying:
\be
\rdot = 0 \; \to \; r = r_0 = \mathrm{const} \quad ,\quad A(r_0) = 0 \; .
\lab{horizon-standard}
\ee
Eq.\rf{horizon-standard} tells us that consistency of \textsl{LL-brane} dynamics in 
a spherically symmetric gravitational background of codimension one requires the 
latter to possess a horizon (at some $r = r_0$), which is automatically occupied 
by the \textsl{LL-brane} (``horizon straddling'' according to the
terminology of Ref.\ct{Barrabes-Israel}). Similar property -- 
``horizon straddling'', has been found also for \textsl{LL-branes} moving in
rotating axially symmetric (Kerr or Kerr-Newman) and rotating cylindrically
symmetric black hole backgrounds \ct{our-WH,rot-WH}.

Next, the Maxwell coupling of the \textsl{LL-brane} produces via 
Eq.\rf{Maxwell-eqs} static Coulomb field in the outer region beyond the
horizon (for $r>r_0$). Namely, inserting in Eq.\rf{Maxwell-eqs}
the embedding ansatz \rf{X-embed} together with \rf{horizon-standard} 
and accounting for \rf{gauge-fix}--\rf{gamma-eqs-0} we obtain;
\be
\pa_r\( C^{p/2}(r)\cF_{vr} (r)\) - 
q\,\frac{\sqrt{p! F^2}}{(2a_0)^{p/2}} C^{p/2}(r_0) \d (r-r_0) = 0 \; ,
\lab{Maxwell-0}
\ee
which yields for the Maxwell field-strength:
\be
\cF_{vr} (r) = \(\frac{C(r_0)}{C(r)}\)^{p/2} 
\frac{q\sqrt{p! F^2}}{(2a_0)^{p/2}}\th(r-r_0) \; .
\lab{F-Maxwell}
\ee

Using again the embedding ansatz \rf{X-embed} together with \rf{horizon-standard}
as well as \rf{gauge-fix}--\rf{gamma-eqs-0}, the Kalb-Ramond equations of motion
\rf{F-KR-eqs} reduce to:
\br
\pa_r \cF + \b \d (r-r_0) = 0 \quad \to \quad 
\cF = \cF_{(+)} \th (r-r_0) + \cF_{(-)} \th (r_0 -r)
\lab{F-KR-0}\\
\cF_{(\pm)} = \mathrm{const} \quad ,\quad \cF_{(-)} - \cF_{(+)} = \b
\lab{F-jump}
\er
Therefore, a space-time varying non-negative cosmological constant is dynamically 
generated in both exterior and interior regions w.r.t. the horizon at $r=r_0$
(cf. Eq.\rf{T-KR}):
\be
\L_{(\pm)} = 4\pi \cF^2_{(\pm)} \; .
\lab{cosmolog-const}
\ee

Finally, it remains to consider the second order (w.r.t. proper time derivative) 
$X^\m$ equations of motion \rf{X-eqs}. Upon inserting the embedding ansatz
\rf{X-embed} together with \rf{horizon-standard} and taking into account
\rf{gamma-eqs-0}, \rf{F-Maxwell} and \rf{F-jump}, we find that the only non-trivial
equations is for $\m=v$. Before proceeding let us note that the ``force''
terms in the $X^\m$ equations of motion \rf{X-eqs} (the geodesic ones
containing the Christoffel connection coefficients as well as those coming
from the \textsl{LL-brane} coupling to the bulk Maxwell and Kalb-Ramond
gauge fields) contain discontinuities across the horizon occupied by the
\textsl{LL-brane}. The discontinuity problem is resolved following 
the approach in Ref.\ct{Israel-66} (see also the regularization
approach in Ref.\ct{BGG}, Appendix A) by taking mean values of the ``force''
terms across the discontinuity at $r=r_0$. Thus, we obtain from Eq.\rf{X-eqs}
with $\m=v$:
\br
\pa_\t \chi + \chi \llb \frac{1}{4}\(\pa_r A_{(+)} + \pa_r A_{(-)}\) 
+ \h p a_0 \(\pa_r \ln C_{(+)} + \pa_r \ln C_{(-)}\)\rrb_{r=r_0}
\nonu \\
+ \h \Bigl\lb - q^2\frac{p! F^2}{(2a_0)^{p/2}}
+ \b (2a_0)^{p/2} \(\cF_{(-)} + \cF_{(+)}\)\Bigr\rb = 0
\lab{X-0-eq} \; .
\er


\section{Asymmetric Wormhole Solution}

The \textsl{LL-brane} energy-momentum tensor \rf{T-brane} on the r.h.s. of
the Einstein equations of motion \rf{Einstein-eqs}, upon inserting the 
expressions for $X^\m (\s)$ from \rf{X-embed} and \rf{horizon-standard}, 
and taking into account the gauge-fixing conditions \rf{gauge-fix} and the
ansatz \rf{F-ansatz}, acquires the form:
\be
T_{(brane)}^{\m\n} = S^{\m\n}\,\d (r-r_0)
\lab{T-S-0}
\ee
with surface energy-momentum tensor:
\be
S^{\m\n} \equiv \frac{\chi}{(2a_0)^{p/2}}\,
\llb \pa_\t X^\m \pa_\t X^\n - 2a_0 G^{ij} \pa_i X^\m \pa_j X^\n 
\rrb_{v=\t,\,r=r_0,\,\th^i =\s^i} \; .
\lab{T-S-brane}
\ee
Here $a_0$ is the integration constant parameter appearing in the 
\textsl{LL-brane} dynamics \rf{a0-const} and $G_{ij} = C(r) h_{ij}(\th)$.
For the non-zero components of $S_{\m\n}$ (with lower indices) and its trace we find:
\be
S_{rr} = \frac{\chi}{(2a_0)^{p/2}} \quad ,\quad 
S_{ij} = - \frac{\chi}{(2a_0)^{p/2-1}} G_{ij} \quad ,\quad
S^\l_\l = - \frac{p\chi}{(2a_0)^{p/2-1}} 
\lab{S-comp}
\ee
The solution of the other bulk space-time equations of motion (the Maxwell
\rf{Maxwell-eqs} and Kalb-Ramond \rf{F-KR-eqs}) with spherically symmetric
geometry have already been given in the previous Section, see
Eqs.\rf{Maxwell-0}--\rf{F-jump}.

For the sake of simplicity we will consider in what follows the case of 
$D=4$-dimensional bulk space-time and, correspondingly, $p=2$ for the 
\textsl{LL-brane}. The generalization to arbitrary $D$ is straightforward.
For further simplification of the numerical constant factors we will choose
the following specific (``wrong-sign'' Maxwell) form for the Lagrangian of the 
auxiliary non-dynamical world-volume gauge field (cf. Eqs.\rf{p-rank}--\rf{F2-id}): 
\be
L(F^2)=\frac{1}{4}F^2 \quad \to \quad  a_0 = M \; ,
\lab{L-eq-0}
\ee
where again $a_0$ is the constant defined in \rf{a0-const} and
$M$ denotes the original integration constant in Eqs.\rf{phi-eqs}. 

We will show that there exists an {\em asymmetric wormhole} solution of the
Einstein equations of motion \rf{Einstein-eqs} with \textsl{LL-brane}
energy-momentum tensor on the r.h.s. given by \rf{T-S-brane}--\rf{S-comp} --
systematically derived from the reparametrization invariant \textsl{LL-brane}
world-volume action \rf{LL-action}, which describes two ``universes'' with
different spherically symmetric geometries \rf{EF-metric} 
matched along a \textsl{LL-brane}. More specifically, this solution
describes an overall space-time manifold containing two separate
spherically symmetric space-time regions:

(i) a ``left universe'' consisting of
the exterior region of Schwarzschild-de-Sitter space-time above the {\em inner}
(Schwarzschild-type) horizon, \textsl{i.e.}, with metric \rf{EF-metric} where
$C(r) = r^2$ and:
\br
A(r) \equiv A_{(-)}(r) = 1 - \frac{2m_1}{r} - K r^2 
\qquad \mathrm{for}\; r > r_0 \; ,
\lab{left-univ} \\
A_{(-)}(r_0) = 0  \quad ,\quad \pa_r A_{(-)}\!\!\bv_{r=r_0} >0 \; ;
\lab{left-univ-a}
\er

(ii) a ``right universe'' comprising the
exterior Reissner-Nordstr{\"o}m black hole region beyond the {\em outer}
Reissner-Nordstr{\"o}m horizon with metric \rf{EF-metric} where $C(r) = r^2$ and:
\br
A(r) \equiv A_{(+)}(r) = 1 - \frac{2m_2}{r} + \frac{Q^2}{r^2}
\qquad \mathrm{for}\;  r > r_0 \; ,
\lab{right-univ} \\
A_{(+)}(r_0) = 0  \quad ,\quad \pa_r A_{(+)}\!\!\bv_{r=r_0} >0 \; .
\lab{right-univ-a}
\er

The ``throat'' connecting the above two ``universes'' with metrics \rf{left-univ} and
\rf{right-univ} is the lightlike world-volume hypersurface of the
\textsl{LL-brane} which is located on the common horizon ($r=r_0$) of both
``universes'' -- a Schwarzschild-type horizon from the ``left universe'' side
\rf{left-univ-a} and an outer Reissner-Nordstr{\"o}m horizon from the 
``right universe'' side \rf{right-univ-a}.
As already pointed out above (Eqs.\rf{r-const}--\rf{horizon-standard}), the common
horizon at $r=r_0$ is automatically occupied by the \textsl{LL-brane} 
(``horizon straddling'') as a result of its world-volume dynamics. 

The asymmetric wormhole solution under consideration is explicitly given in
Eddington-Finkelstein-type coordinates as:
\be
ds^2 = - {\wti A} (\eta) dv^2 + 2 dv\,d\eta +
{\wti r}^2(\eta) \llb d\th^2 + \sin^2\th d\vp^2\rrb \; ,
\lab{our-EF-metric}
\ee
where the original radial coordinate $r$ ($r>0$) is replaced by $\eta$
($-\infty <\eta <\infty$) upon substituting: 
\be
r \;\;\to\;\; {\wti r}(\eta) = r_0 + |\eta|
\lab{r-to-eta}
\ee
with $r_0$ -- the common horizon of
\rf{left-univ} and \rf{right-univ} as follows:
\br
{\wti A} (\eta) \equiv A_{(-)} \bigl({\wti r}(\eta)\bigr) =
1 - \frac{2m_1}{{\wti r}(\eta)}  - K {\wti r}^2(\eta) \quad ,\;\; 
K \equiv \frac{4\pi}{3}\b^2 \;\; ,\;\; \mathrm{for}\; \eta <0 \; ,
\lab{left-univ-1}
\\
{\wti A} (\eta) \equiv A_{(+)} \bigl({\wti r}(\eta)\bigr) =
1 - \frac{2m_2}{{\wti r}(\eta)} + \frac{Q^2}{{\wti r}^2(\eta)} \quad , \;\; 
Q^2 \equiv \frac{8\pi}{a_0} q^2\, r_0^4 \;\;, \;\; \mathrm{for}\; \eta >0 \; .
\lab{right-univ-1}
\er
The new radial-like coordinate $\eta$ describes a continuous interpolation
between the left ($\eta<0$) and the right ($\eta>0$) ``universes'' through 
the ``throat'' at $\eta=0$.
As shown in Eqs.\rf{F-KR-0}--\rf{cosmolog-const}, the \textsl{LL-brane} 
through its coupling to the bulk Kalb-Ramond field (cf. \rf{LL-action+EM+KR}
and \rf{F-KR-eqs}) dynamically generates space-time varying non-negative
cosmological constant with a jump across the horizon ($r=r_0$). In the
present case this yields dynamically generated de Sitter parameter
$K = \frac{1}{3} \L_{(-)} \equiv \frac{4\pi}{3} \cF_{(-)}^2 = \frac{4\pi}{3}\b^2$ 
in the ``left universe'' as given in \rf{left-univ-1}, whereas $\cF_{(+)}=0$. 
On the other hand, the surface charge density $q$ of the \textsl{LL-brane} 
(cf. \rf{Maxwell-eqs} and \rf{Maxwell-0}) explicitly determines the non-zero 
Coulomb field-strength in the ``right universe'' with $Q^2$ as given in
\rf{right-univ-1}.

Now, substituting the metric \rf{our-EF-metric} with
\rf{left-univ-1}--\rf{right-univ-1} into the Einstein equations \rf{Einstein-eqs} 
and taking into account that outside the ``throat'' ($\eta=0$)
it obviously solves the ``vacuum'' equations (with the \textsl{LL-brane}
absent), the only non-trivial $\d$-function contribution on the l.h.s. of
\rf{Einstein-eqs} arises because of non-smoothness of the metric 
\rf{our-EF-metric} with \rf{left-univ-1}--\rf{right-univ-1}
at $\eta=0$ (it is continuous but not differentiable there). Thus, inserting 
the world-volume Lagrangian-derived expression \rf{T-S-0}--\rf{T-S-brane} for the 
\textsl{LL-brane} 
stress-energy tensor on the r.h.s. of Einstein Eqs.\rf{Einstein-eqs} yields 
(for $D=4, p=2$) two relations matching the coefficients in front of $\d (\eta)$:
\be
\llb\pa_\eta {\wti A}_{(+)} - \pa_\eta {\wti A}_{(-)}\rrb_{\eta=0} =
- 16\pi \chi \quad , \quad r_0 = - \frac{a_0}{\pi \chi} \; ,
\lab{match-1}
\ee
The second relation \rf{match-1} implies that the dynamical \textsl{LL-brane} tension 
$\chi$ must be constant (independent of the \textsl{LL-brane} proper time
$\t$) and it must be {\em negative}. Substituting the explicit form 
\rf{left-univ-1}--\rf{right-univ-1} of ${\wti A}_{\pm}$ in \rf{match-1} gives the 
following expressions for the mass parameters:
\be
m_1 = \frac{a_0}{2\pi |\chi|}\Bigl( 1 - \frac{4a_0^2 \b^2}{3\pi\chi^2}\Bigr)
\quad ,\quad
m_2 = \frac{a_0}{2\pi |\chi|}\Bigl( 1 + \frac{8a_0 q^2}{\pi\chi^2}\Bigr) \; ,
\lab{m1-m2}
\ee as well as the relation between $a_0$ and $\chi$:
\be
\chi^2 = \frac{2a_0\,(2q^2 + a_0 \b^2)}{\pi (1-8a_0)} \; .
\lab{chi-a0}
\ee
Next, the \textsl{LL-brane} equation of motion \rf{X-0-eq}, where we set
$\chi=\mathrm{const}$, yields in the
case under consideration a third matching relation at the ``throat'':
\be
\frac{|\chi|}{4}\llb \pa_\eta {\wti A}_{(+)} + \pa_\eta {\wti A}_{(-)}\rrb_{\eta=0} +
2q^2 - a_0\b^2 = 0 \; ,
\lab{match-2}
\ee
which upon using \rf{m1-m2}--\rf{chi-a0} reduces to the second relation \rf{match-1},
\textsl{i.e.}, Eq.\rf{match-2} does not carry any new information. 

From relations \rf{m1-m2}--\rf{chi-a0} we conclude that all physical parameters of
the asymmetric wormhole \rf{left-univ-1}--\rf{right-univ-1} are explicitly determined
by the two free parameters $(q,\b)$ -- the surface electric charge density $q$ and 
the Kalb-Ramond charge $\b$ of the \textsl{LL-brane}.

It remains to check that:
\be
\pa_\eta {\wti A}\!\!\bv_{\eta \to +0} \equiv \pa_r A_{(+)}\!\!\bv_{r=r_0} >0 
\quad ,\quad
- \pa_\eta {\wti A}\!\!\bv_{\eta \to -0} \equiv \pa_r A_{(-)}\!\!\bv_{r=r_0} >0 \; ,
\lab{A-inequal}
\ee
\textsl{i.e.}, the ``throat'' at $\eta=0$ must be the outer Reissner-Nordstr{\"o}m 
horizon from the point of view of the ``right'' Reissner-Nordstr{\"o}m 
``universe'' ($\eta > 0$, cf. \rf{right-univ-a}) and simultaneously be the inner
Schwarzschild-type horizon from the point of view of the ``left''
Schwarzschild-de-Sitter ``universe''($\eta < 0$, cf. \rf{left-univ-a}).
From \rf{m1-m2}--\rf{chi-a0} we find:
\br
\pa_\eta {\wti A}\!\!\bv_{\eta \to +0} =
\frac{\pi\,|\chi|}{a_0 (2q^2 + a_0\b^2)} \llb 2q^2 (16a_0 -1) + a_0 \b^2\rrb >0
\lab{der-A-1} \\
- \pa_\eta {\wti A}\!\!\bv_{\eta \to -0} =
\frac{\pi\,|\chi|}{a_0 (2q^2 + a_0\b^2)} \llb 2q^2 + a_0 \b^2 (16a_0 -1)\rrb 
> 0 \; .
\lab{der-A-2}
\er
Inequalities \rf{der-A-1}--\rf{der-A-2} together with Eq.\rf{chi-a0} imply the
following restriction on the integration constant $a_0$ from the
\textsl{LL-brane} dynamics \rf{a0-const}:
\be
1/16 < a_0 < 1/8 \; .
\lab{a0-inequal}
\ee

In complete analogy one can construct another asymmetric wormhole solution where
the \textsl{LL-brane} connects the ``left'' universe, which is now the exterior region of 
the standard Schwarzschild space-time ($r > r_0 = 2m_1$), with the ``right'' universe,
which is the exterior region of the Reissner-Nordstr{\"o}m-de-Sitter space-time
($r > r_0$) beyond the outer Reissner-Nordstr{\"o}m horizon $r=r_0$:
\br
{\wti A} (\eta) \equiv A_{(-)} \bigl({\wti r}(\eta)\bigr) =
1 - \frac{2m_1}{{\wti r}(\eta)} \quad \mathrm{for}\; \eta <0 \; ,
\lab{left-univ-2} \\
{\wti r}(\eta) = r_0 - \eta \quad ,\;\; r_0 = 2m_1 \;\;,
\nonu
\er
\br
{\wti A} (\eta) \equiv A_{(+)} \bigl({\wti r}(\eta)\bigr) =
1 - \frac{2m_2}{{\wti r}(\eta)} + \frac{Q^2}{{\wti r}^2(\eta)} - K {\wti r}^2(\eta)
\quad \mathrm{for}\; \eta >0 \; ,
\lab{right-univ-2} \\
{\wti r}(\eta) = r_0 + \eta \quad ,\;\;
Q^2 \equiv \frac{8\pi}{a_0} q^2\, r_0^4 \quad , \quad
K \equiv \frac{4\pi}{3}\b^2 \; .
\nonu
\er
In this case the physical parameters of the asymmetric wormhole read:
\be
m_1 = \frac{r_0}{2} = \frac{a_0}{2\pi\chi} \quad ,\quad
m_2 = \frac{a_0}{2\pi\chi} \Bigl\lb 1 + \frac{2(1-8a_0)}{2q^2 + a_0\b^2}
\Bigl( 2q^2 -\frac{1}{3}a_0\b^2\Bigr)\Bigr\rb \; ,
\lab{}
\ee
for $\chi^2$ we get the same relation \rf{chi-a0} and the same restriction
\rf{a0-inequal} holds for the \textsl{LL-brane} integration constant $a_0$.

\section{A Note on Einstein-Rosen ``Bridge''}

In the simple special case $(q=0,\b=0)$ the {\em asymmetric} wormhole solution
\rf{our-EF-metric}--\rf{right-univ-1} reduces to a {\em symmetric} wormhole solution: 
\br
ds^2 = - {\wti A} (\eta) dv^2 + 2 dv\,d\eta +
{\wti r}^2(\eta) \llb d\th^2 + \sin^2\th d\vp^2\rrb \; ,
\nonu \\
{\wti A} (\eta) = 1 - \frac{2m}{{\wti r}(\eta)} \quad ,\quad
{\wti r}(\eta) = 2m + |\eta| \; .
\lab{our-ER}
\er
The above metric describes two identical copies of Schwarzschild {\em exterior}
space-time region ($r > 2m$), 
which correspond to
$\eta >0$ and $\eta <0$, respectively, and which are ``glued'' together at the
horizon $\eta = 0$ (\textsl{i.e.}, $r=2m$) occupied by the \textsl{LL-brane},
where the latter serves as a throat of the overall wormhole solution.
This is precisely the space-time manifold of the Einstein-Rosen ``bridge''
solution in its original formulation \ct{einstein-rosen} in terms of
Eddington-Finkelstein coordinates. An important consequence of
the present construction is that the Einstein-Rosen ``bridge'' wormhole {\em does
not} satisfy the vacuum Einstein equations since it needs the presence of a
non-trivial matter stress-energy tensor on the r.h.s. \rf{Einstein-eqs} which turns 
out to be the stress-energy tensor of a \textsl{LL-brane} \rf{T-S-0}--\rf{T-S-brane} 
self-consistently derived
from a well-defined reparametrization-invariant world-volume \textsl{LL-brane} 
Lagrangian \rf{LL-action}.

To make the connection with the original formulation \ct{einstein-rosen} of the 
Einstein-Rosen ``bridge'' more explicit let us recall that Einstein and
Rosen start from the standard Schwarzschild metric:
\be
ds^2 = - A(r) dt^2 + A^{-1}(r) dr^2 + r^2 \llb d\th^2 + \sin^2\th\, d\vp^2\rrb
\quad , \;\; A(r) = 1 - \frac{2m}{r}
\lab{Schw}
\ee
and introduce new radial-like coordinate $u$ by defining $u^2=r-2m$, so that 
the metric \rf{Schw} becomes:
\be
ds^2 = - \frac{u^2}{u^2 + 2m} dt^2 + 4 (u^2 + 2m)du^2 +
(u^2 + 2m)^2 \llb d\th^2 + \sin^2 \th \,d\vp^2\rrb \; .
\lab{E-R-metric}
\ee
Then Einstein and Rosen take two identical copies of the exterior Schwarzschild
space-time region ($r>2m$) by letting the new coordinate $u$ to vary between
$-\infty$ and $+\infty$
(\textsl{i.e.}, we have the same $r\geq 2m$ for $\pm u$). The two
Schwarzschild exterior space-time regions must be matched at the horizon $u=0$
(the wormhole ``throat'').

Let us examine whether the original Einstein-Rosen solution satisfy the
vacuum Einstein equations everywhere. To this end let us consider the Levi-Civita
identity (see \textsl{e.g.} \ct{frankel}):
\be
R^0_0 = - \frac{1}{\sqrt{-g_{00}}} \nabla^2 \(\sqrt{-g_{00}}\)
\lab{levi-civita-id}
\ee
valid for any metric of the form
$ds^2 = g_{00} (r) (dt)^2 + h_{ij}(r,\th,\vp) dx^i dx^j$ and where $\nabla^2$
is the three-dimensional Laplace-Beltrami operator
$\nabla^2 =\frac{1}{\sqrt{h}}\partder{}{x^i}\(\sqrt{h}\,h^{ij}\partder{}{x^j}\)$.
The Einstein-Rosen metric \rf{E-R-metric} solves $R^0_0 = 0$ for all $u\neq 0$.
However, since $\sqrt{-g_{00}} \sim |u|$ as $u \to 0$ and since
$\frac{\pa^2}{{\pa u}^2} |u| = 2 \d (u)$, Eq.\rf{levi-civita-id} tells us that:
\be
R^0_0 \sim \frac{1}{|u|} \d (u) \sim \d (u^2) \; ,
\lab{ricci-delta}
\ee
and similarly for the scalar curvature $R \sim \frac{1}{|u|} \d (u) \sim \d (u^2)$.
From \rf{ricci-delta} we conclude that:

(i) The non-vanishing r.h.s. of \rf{ricci-delta} exhibits the explicit presence of
some lightlike matter source on the throat -- an observation which is missing in
the original formulation \ct{einstein-rosen} of the Einstein-Rosen ``bridge''.
In fact, the problem with the metric \rf{E-R-metric} satisfying the vacuum Einstein
equations at $u=0$ has been noticed in ref.\ct{einstein-rosen}, where in Eq.(3a)
the authors multiply Ricci tensor by an appropriate power of the determinant $g$
of the metric \rf{E-R-metric} vanishing at $u=0$ so as to enforce
fulfillment of the vacuum Einstein equations everywhere, including at $u=0$.

(ii) The coordinate $u$ in \rf{E-R-metric} is {\em inadequate} 
for description of the original Einstein-Rosen ``bridge'' at the throat due to
the {\em ill-definiteness} as distribution of the r.h.s. in \rf{ricci-delta}.

As we have seen from our construction above, the proper radial-like coordinate for the
Einstein-Rosen ``bridge'' wormhole is $\eta$ which is related to the 
Einstein-Rosen coordinate $u$ via {\em non-smooth} transformation:
\be
u = \mathrm{sign}(\eta)\,\sqrt{|\eta|} \quad,
\quad \mathrm{i.e.}\;\; u^2 = |\eta| \; .
\lab{u-eta}
\ee
Thus, we conclude that solution \rf{our-ER} is the proper self-consistent
formulation of the original Einstein-Rosen ``bridge'' wormhole where the presence of
\textsl{LL-brane} matter source at the ``throat'' plays crucial role for its
well-definiteness.


\section{Conclusions}

I this work we have continued to explore the use of codimension-one 
\textsl{LL-branes} for construction of wormhole solutions of Einstein equations, 
in the present case -- constructing asymmetric wormholes. 
We have strongly emphasized the crucial properties of the dynamics of 
\textsl{LL-branes} interacting with gravity and bulk space-time gauge fields:

(i) The \textsl{LL-brane} automatically locates itself on (one of) the
horizon(s) of the bulk space-time geometry (``horizon straddling''); 

(ii) The \textsl{LL-brane} tension is an additional {\em dynamical degree of 
freedom} unlike the case of standard Nambu-Goto $p$-branes (where it is a given
\textsl{ad hoc} constant), and which might in particular acquire negative values;

(iii) The \textsl{LL-brane} stress-energy tensor provides the appropriate
source term on the r.h.s. of Einstein equations to enable the existence of
consistent non-trivial wormhole solutions; 

(iv) Electrically neutral \textsl{LL-branes} produce {\em symmetric}
wormholes, \textsl{i.e.}, where both left and right ``universes'' are
related via reflection symmetry and where, in particular, the Misner-Wheeler
``charge without charge'' \ct{misner-wheeler} phenomenon is observed;

(v) The wormhole reflection symmetry is broken through the natural couplings
of the \textsl{LL-brane} to bulk space-time gauge fields (Maxwell and
3-index Kalb-Ramond). In this way the \textsl{LL-brane} dynamically generates 
non-zero Coulomb field-strength in the ``right'' universe and non-zero cosmological 
constant either in the ``left'' or in the ``right'' universe which enable the existence 
of {\em asymmetric} (with {\em no} reflection symmetry) wormholes.

We have specifically stressed the crucial role of \textsl{LL-branes} already in 
the case of the ``mother of all wormholes'' -- the classic Einstein-Rosen ``bridge''
manifold \ct{einstein-rosen}.

Finally, let us mention the crucial role of \textsl{LL-branes} in
constructing non-trivial examples of {\em non-singular} black holes,
\textsl{i.e.}, solutions of Einstein equations with black hole type geometry
in the bulk space-time, in particular possessing horizons, but with {\em no}
space-time singularities in the center of the geometry. For further details
we refer to \ct{reg-BH}, where a solution of the Einstein-Maxwell-Kalb-Ramond
system coupled to a charged \textsl{LL-brane} has been obtained describing a 
{\rm regular} black hole. The space-time manifold of the latter consists of 
de Sitter interior region and exterior Reissner-Nordstr{\"o}m region glued
together along their common horizon (it is the {\em inner} horizon from the 
Reissner-Nordstr{\"o}m side).

%


\section*{Acknowledgments}
E.N. and S.P. are supported by Bulgarian NSF grant \textsl{DO 02-257}.
Also, all of us acknowledge support of our collaboration through the exchange
agreement between the Ben-Gurion University of the Negev (Beer-Sheva, Israel) and
the Bulgarian Academy of Sciences.


\end{document}